\documentclass[english, letterpaper, 10 pt, conference]{ieeeconf}
\usepackage[T1]{fontenc}
\usepackage[latin9]{inputenc}
\usepackage{amsmath}
\usepackage{cite}
\usepackage[caption=false]{subfig}
\usepackage{dsfont}
\usepackage{bm}
\usepackage{url}

\IEEEoverridecommandlockouts
\makeatletter

\usepackage{tikz}
\usepackage{tikzscale}
\usepackage{amssymb}

\usepackage{enumerate}   
\usepackage{graphicx} 
\usepackage[binary-units=true]{siunitx}
\usepackage{pgfplots}
\pgfplotsset{compat = 1.12}
\usepackage{cleveref}
\usepackage[ruled,vlined]{algorithm2e}
\usepackage{booktabs}
\usetikzlibrary{external}
\pgfplotsset{select coords between index/.style 2 args={
		x filter/.code={
			\ifnum\coordindex<#1\fi
			\ifnum\coordindex>#2\fi
		}
}}
\usepgfplotslibrary{groupplots}
\usetikzlibrary{matrix}

\newcounter{tagnumb}
\newcommand{\bx}{\xi}
\newcommand{\bu}{\mathbf u}

\crefname{algocf}{alg.}{algs.}
\Crefname{algocf}{Algorithm}{Algorithms}

\newcommand{\includegtikz}[2][]{ 
	\tikzsetnextfilename{#2}
	\includegraphics[#1]{figures/tikz/#2} }

\makeatother

\usepackage{babel}

\definecolor{col1}{rgb:Hsb}{210,1,1}
\definecolor{col2}{rgb:Hsb}{220,1,0.8}
\definecolor{col3}{rgb:Hsb}{230,1,0.6}
\definecolor{col4}{rgb:Hsb}{240,1,0.4}
\definecolor{col5}{rgb:Hsb}{250,1,0.2}

\title{\LARGE A Simple Dynamic Model for Aggressive, Near-Limits Trajectory Planning}

\author{Florent Altch\'e$^{2,1}$, Philip Polack$^{1}$ and Arnaud de La Fortelle$^{1}$
\thanks{$^{1}$ MINES ParisTech, PSL Research University, Centre for robotics, 60 Bd St Michel 75006 Paris, France {\tt\small [florent.altche, philip.polack, arnaud.de\_la\_fortelle] @mines-paristech.fr}}
\thanks{$^{2}$ \'Ecole des Ponts ParisTech, Cit\'e Descartes, 6-8 Av Blaise Pascal, 77455 Champs-sur-Marne, France}%
}

\begin{document}

\maketitle
\thispagestyle{empty}
\pagestyle{empty}

\begin{abstract} 
	In normal on-road situations, autonomous vehicles will be expected to have smooth trajectories with relatively little demand on the vehicle dynamics to ensure passenger comfort and driving safety. However, the occurrence of unexpected events may require vehicles to perform aggressive maneuvers, near the limits of their dynamic capacities. In order to ensure the occupant's safety in these situations, the ability to plan controllable but near-limits trajectories will be of very high importance. One of the main issues in planning aggressive maneuvers lies in the high complexity of the vehicle dynamics near the handling limits, which effectively makes state-of-the-art methods such as Model Predictive Control difficult to use. This article studies a highly precise model of the vehicle body to derive a simpler, constrained second-order integrator dynamic model which remains precise even near the handling limits of the vehicle. Preliminary simulation results indicate that our model provides better accuracy without increasing computation time compared to a more classical kinematic bicycle model. The proposed model can find applications for contingency planning, which may require aggressive maneuvers, or for trajectory planning at high speed, for instance in racing applications.
\end{abstract}

\section{Introduction} 
Planning safe and efficient trajectories remains an important challenge for autonomous driving, in particular when approaching the limits of handling of the vehicle. Arguably, most driving situations do not require pushing the vehicle to its limits; however, some situations may require the ability to plan ``aggressive'' maneuvers to guarantee the safety of the vehicle and its occupants, for instance when driving at high speed or in low adherence conditions. The ability to plan aggressive maneuvers can also be beneficial to compute contingency trajectories in parallel with a comfortable ``reference'' one. For instance, aborting an overtaking maneuver due to new perception data involves combined braking and steering, which can result in loss of adherence at high speed. To ensure that contingency trajectories are feasible, such planners should be able to precisely take into account the vehicle's dynamic limitations.

One of the main difficulties of aggressive trajectory planning lies in the heavy nonlinearities of the vehicle dynamics when close to its handling limits. These nonlinearities arise from various phenomenons, and are often difficult to take into account in a trajectory planner. For this reason, the road-tire forces and the variation of the vertical forces on each wheel (load transfer) are often ignored or extremely simplified at the planning stage. In the existing literature, many authors (see, \textit{e.g.}~\cite{Abbas2014,Cardoso2016}) simply rely on a kinematic modeling of the vehicle as a bicycle, which can result in planned trajectories which are infeasible in practice, or inefficient because safety margins have been chosen too large, leading to overcautious driving.

Some authors have considered more realistic vehicle dynamics for aggressive maneuvers or when driving on slippery roads, for instance in the presence of snow~\cite{Borrelli2005}. For such demanding scenarios, a majority of references use Model Predictive Control (MPC) techniques with a more precise vehicle model, allowing to simultaneously plan a trajectory and compute the corresponding feasible control. In this case, most authors consider a dynamic bicycle model~\cite{Falcone2007,Park2009,Gao2010,Liu2014,Ji2016} with various levels of complexity in the modeling of the tire forces.

The main limitation of finer vehicle models which include wheel dynamics is that the wheel velocities generally have much shorter characteristic times (around \SI{1}{\milli\second}~\cite{Altrock1994}) than the vehicle's dynamics (typically \SI{100}{\milli\second}). As a result, models taking wheel dynamics into account require a very short integration time step in MPC formulations, which greatly reduces the planning horizon that can be considered in real-time computation. The main contribution of this article is an alternative approach to take into account finer information about the vehicle dynamics while remaining computationally tractable over a planning horizon of a few seconds. Instead of directly using the (highly complex) dynamic equations of the vehicle during online solving, we first compute offline the set of feasible longitudinal, lateral and angular accelerations for various initial states of the vehicle. We then propose a convex approximation of this feasible region, which allows reformulating the vehicle dynamics using a carefully constrained second-order integrator model. The reduced complexity of this model makes it easy to implement as an MPC planner, and allows using longer time steps for numerical optimization.

Note that some authors have already studied approximations for the set of reachable accelerations. A commonly used model is the so-called ``friction circle''~\cite{Goh2016} (or ellipse~\cite{Cowlagi2012,Choi2014}), in which this set is approximated by a simple Coulomb modeling of the friction forces. However, this approximation is generally used to account for slip at the wheel level, and thus requires modeling wheel dynamics. Another possible approach is to directly measure actual acceleration data on a test vehicle; such results have been summarized in a so-called ``g-g diagram'' in~\cite{Funke2012}, but these experiments are difficult and costly to perform. In this article, we use a simulation-based approach to compute synthetic feasibility envelopes for the vehicle. Interestingly, our results show that these envelopes (computed for the whole vehicle) are actually closely related to the g-g diagram used in~\cite{Funke2012,Zhan2017}.



The rest of this article is structured as follows: in \Cref{sec:vehicle-model}, we present a 9 degrees of freedom dynamic model of the vehicle's body, that we will use throughout the rest of this article. In \Cref{sec:computation-feasible}, we describe an offline method to compute the sets of feasible accelerations for the vehicle, and we derive a constrained second-order integrator dynamic model in \Cref{sec:2d-integrator}. In \Cref{sec:numerical-sim}, we provide preliminary simulation results, comparing the proposed model with a more classical kinematic bicycle one. Finally, \Cref{sec:conclusion} concludes the study.

\section{Vehicle model\label{sec:vehicle-model}}
\begin{figure}[h!]
	\centering
	\includegraphics[scale=0.4]{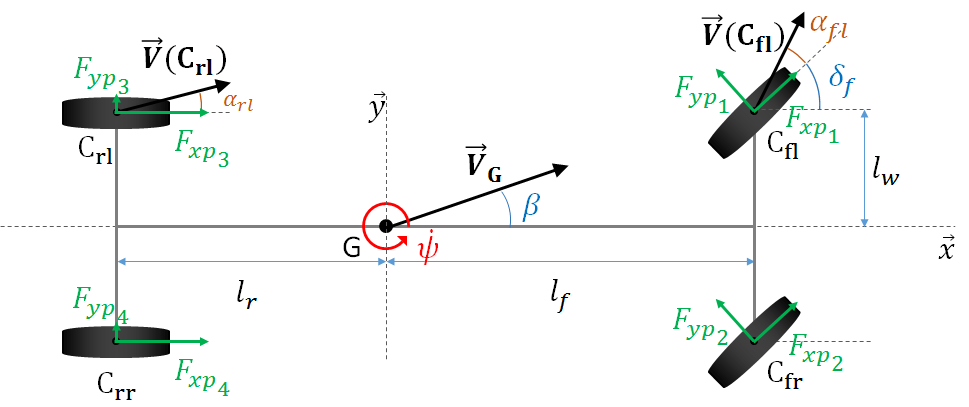}       
	\caption{Simulation model of the vehicle in the $(x,y)$ plane}
	\label{fig:carSim}
\end{figure}

In this section, we describe a 9 degrees of freedom vehicle body model that we will use throughout this article. Alongside with the usual 2D state $[X, Y, \psi]$ (with $\psi$ the yaw rotation) of the vehicle, the model takes into account its roll and pitch movements, the dynamics of the wheels and the coupling of longitudinal and lateral slips of the tires. Being a chassis model, it does not take into account the dynamics of the car engine or brakes. The control inputs of the vehicle are the torque $T_i$ applied to each wheel $i$ and the steering angle of the front wheels, $\delta$. In this article, we use uppercase letters (\textit{e.g.}, $X$, $Y$) to denote coordinates in the ground (global) frame, and lowercase letters for coordinates in the vehicle (local) frame; the $x$ coordinate in the local frame corresponds to the longitudinal component. The notations are given in \Cref{tab:notations} and illustrated in \Cref{fig:carSim}.


\begin{table}[h]
	\caption{Notations}
	\label{tab:notations}
	\begin{tabular}{p{1.45cm} p{6.35cm}}
		\toprule
		$X$, $Y$, $Z$ & Position of the vehicle's CoM (ground frame) \\
		$\theta$, $\phi$, $\psi$ & Roll, pitch and yaw angles of the car body \\
		$V_x$, $V_y$ & Longitudinal and lat. vehicle speed (vehicle frame) \\
		$V_{xw_i}$ & Longitudinal speed of wheel $i$ (wheel frame)\\
		$\omega_i$ & Angular velocity of wheel $i$ \\
		$\zeta_i$ & Displacement of suspension $i$ \\
		$\delta$ & Steering angle of the front wheels \\
		$T_{\omega_i}$ & Total torque applied to wheel $i$\\
		$F_{xw_i}$, $F_{yw_i}$  & Longitudinal and lateral forces on wheel $i$ (wheel frame)\\
		$F_{x_i}$, $F_{y_i}$ & Longitudinal and lat. forces on wheel $i$ (vehicle frame)\\
		$F_{z_i}$ & Normal ground force on wheel $i$\\
		$F_{aero}$ & Air drag force on the vehicle \\
		$M_T$ & Total mass of the vehicle\\
		$I_x$, $I_y$, $I_z$ & Roll, pitch and yaw inertia of the vehicle \\
		$I_{r_i}$ & Inertia of wheel $i$ around its axis \\
		$l_f$, $l_r$ & Distance between the front/rear axle and the CoM\\
		$l_w$ & Half-track of the vehicle\\ 
		$r_{w}$ & Effective radius of the wheels \\ 
		$k_s$, $d_s$ & Suspensions stiffness and damping \\
		\bottomrule
	\end{tabular}
\end{table}

In what follows, we assume that the body of the vehicle rotates around its center of mass, and that the aerodynamic forces do not create a moment on the vehicle. Moreover, we assume that the road remains horizontal, and any slope or banking angle is neglected; this assumption could be relaxed using a slightly more complex vehicle model. Under these hypotheses, the dynamics of the vehicle's center of mass are written as:
\begin{subequations}\label{eq:globalframe}
	\begin{align}
	\dot{X} =\ & V_x \cos\psi - V_y \sin\psi\\
	\dot{Y} =\ & V_x \sin\psi + V_y \cos\psi	\\
	\dot{V}_x =\ & \dot{\psi} V_y + \frac{1}{M_T}\sum_{i=1}^4 F_{x_i} - F_{aero}\\
	\dot{V}_y =\ & - \dot{\psi} V_x + \frac{1}{M_T} \sum_{i=1}^4 F_{y_i},
	\end{align}
\end{subequations}
where $F_{x_i}$ and $F_{y_i}$ are respectively the longitudinal and lateral tire forces generated on wheel $i$, expressed in the local vehicle frame $(x,y)$. The yaw, roll and pitch motions of the car body are computed as:
\begin{subequations}\label{eq:rollpitch}
\begin{align}
	I_z\ddot{\psi} = l_f (F_{y_1} + F_{y_2})  - l_r (F_{y_3} + F_{y_4}) &  \nonumber\\
	 +\, l_w (F_{x_2}+F_{x_4}-\, & F_{x_1}-F_{x_3}) \\
	I_x\ddot{\theta} = l_w (F_{z_1}+F_{z_3}-F_{z_2}-F_{z_4})\, +\, & Z \sum_{i=1}^4 F_{y_i}\\
	I_y\ddot{\phi} = l_r (F_{z_3} + F_{z_4}) -l_f (F_{z_1} +F_{z_2})\, -\, & Z \sum_{i=1}^4 F_{x_i}
\end{align}
\end{subequations}
where $F_{z_i} = - k_s \zeta_{i}(\theta, \phi) - d_s \dot{(\zeta_{i})}(\theta, \phi)$, with $\zeta_i(\theta,\phi)$ the displacement of suspension $i$ for the given roll and pitch angles of the car body. The variation of $F_z$ models the impact of load transfer between tires. Finally, the dynamics of each wheel $i$ can be written as
\begin{align} 
	I_r \dot{\omega}_i = T_{{\omega}_i}-r_w F_{xw_i}.
\end{align}

In general, the longitudinal and lateral forces $F_{xw_i}$ and $F_{yw_i}$ depend on the longitudinal slip ratio $\tau_{i}$, the side-slip angle $\alpha_i$, the reactive normal force $F_{z_i}$ and the road friction coefficient $\mu$. The slip ratio of wheel $i$ can be computed as
\begin{equation}\label{eq:slip-ratio}
	\tau_i = \left\{ 
		\begin{array}{l l}
			\frac{r_w \omega_i - V_{xw_i}}{r_w\omega_i} & \text{if } r_w\omega_i \geq V_{xw_i} \\
			\frac{r_w \omega_i - V_{xw_i}}{V_{xw_i}} & \text{otherwise}.
		\end{array} 
		\right.
\end{equation}

The lateral slip angle $\alpha_i$ of tire $i$ is the angle between the wheel orientation and its velocity, and can be expressed as
\begin{align}
\alpha_f &= \delta - \arctan \frac{V_y + l_f \dot{\psi}}{V_x \pm l_w \dot{\psi}} \\
\alpha_r &= - \arctan \frac{V_y - l_r \dot{\psi}}{V_x \pm l_w \dot{\psi}}
\end{align}
where $f$ and $r$ denote the front and rear wheels.

In this article, we use Pacejka's combined slip tire model (equations (4.E1) to (4.E67) in~\cite{pacejka2005tire}), which takes into account the interaction between longitudinal and lateral slips, 
thus encompassing the notion of friction circle~\cite{guntur1980friction}. For clarity purposes, we do not reproduce the complete set of equations here.



%
%

\section{Feasible acceleration sets\label{sec:computation-feasible}}
In theory, it is possible to use the dynamic model presented in \Cref{sec:vehicle-model} inside a Model Predictive Control (MPC) scheme to compute an optimal control in the form of applied engine and braking torques on the wheels, and a steering angle for the front wheels. However, the corresponding optimization problem would involve a highly nonlinear, nonconvex objective function which furthermore is non-differentiable due to the disjunction \eqref{eq:slip-ratio}. In practice, most available solvers seem unable to handle this problem, except for extremely simple situations.

Several ways around this limitation have been proposed in the literature, in order to take into account chassis and tire dynamics in an MPC formulation. In~\cite{Borrelli2005}, the authors use the wheels slip ratios instead of the applied torque as control variables, and assume that a low-level controller can adjust  wheel velocities accordingly. However, the feasible dynamics of the slip ratio have not been studied yet, and the  low-level control proposed by the authors is limited to relatively low slip, remaining in the linear portion of the Pacejka model.

In this article, we only consider the dynamic response of the car body, and in particular we do not precisely model engine response. Instead, we assume that the engine can deliver a torque comprised between \SI{0}{\newton \meter} and $2 T_{max} > 0$, and that the brakes can apply a negative torque between $T_{min} < 0$ and \SI{0}{\newton \meter} on each wheel. The engine torque is equally split between the two front wheels, and braking torques are supposed equal for wheels on a same axle. Note that torque vectoring~\cite{Siampis2015}, in which the accelerating and braking torques are not equally divided between the wheels of an axle, can also be treated using the same method. The steering angle of the front wheels is supposed to be bounded between $\delta_{min} < 0$ and $\delta_{max} > 0$. With these hypotheses, we note $\mathcal U = [T_{min}, T_{max}] \times [T_{min}, 0] \times [\delta_{min}, \delta_{max}]$ the set of admissible controls and $u = [T_f, T_r, \delta] \in \mathcal U$ a control, where $T_f$ is the torque applied on each of the front wheels, $T_r$ the torque on the rear wheels, and $\delta$ the steering angle for the front wheels.

Using the dynamic model of \Cref{sec:vehicle-model} and starting from a known system state $\bx_0$, it is possible to compute future states of the vehicle under a known control input using numerical integration. In this article, we use a fourth order Runge-Kutta integration scheme with a time step duration $\Delta t$ of \SI{1}{\milli \second}, which appears to be sufficient to correctly handle the wheel dynamics. We compute an approximation of the set of feasible accelerations starting from $\bx_0$ as presented in \Cref{alg:sampling}. In the algorithm, the function fitpolynom($st$, $\star$, 2) returns the coefficients of the best fitting polynom of order $2$ for the component $\star$ of $st$, with leading coefficient first. Therefore, the variable \texttt{feas} contains the set of resulting accelerations in the $X$ and $Y$ directions (noted $a_X$ and $a_Y$) as well as the yaw rate acceleration $\ddot \psi$ (noted $a_\psi$), all expressed in the ground coordinates frame. In what follows, we present outputs from \Cref{alg:sampling} for varying conditions. The control bounds are chosen as $T_{min} = \SI{-1500}{\newton \meter}$, $T_{max} = \SI{1250}{\newton \meter}$ and $\delta_{max} = -\delta_{min} = \SI{30}{°}$. 

\begin{algorithm}  \DontPrintSemicolon
	\caption{Sampling of the feasible regions\label{alg:sampling}}
	\KwData{state $\xi_0$, num. of samples $n$, horizon $T$, step $\Delta t$}
	set \texttt{feas} := []\;
	\For{$i = 1\dots n$}{
		randomly choose $u \in \mathcal U$\;
		\For{$k = 1\dots T/\Delta_t$}{
			set $\xi_k :=$ RK4 ($\xi_{k-1}$, $u, \Delta t)$
		}
		set ${st}$ := ($\xi_k$)$_{k=0\dots T/\Delta t} $\;
		set ${p_X}$ := fitpolynom($st$, X, 2)\;
		set ${p_Y}$ := fitpolynom(${st}$, Y, 2)\;
		set $p_{\psi}$ := fitpolynom(${st}$, $\psi$, 2)\;
		append to: \texttt{feas}, $2\cdot [{p_X}(1), {p_Y}(1), p_{\psi}(1)]$
	}
\end{algorithm}

\subsection{Longitudinal velocity}

\begin{figure}[h!]
	\subfloat[Lateral vs. longitudinal\label{fig:feasible-axay}]{
		\includegtikz[width=0.9\columnwidth,height=6cm]{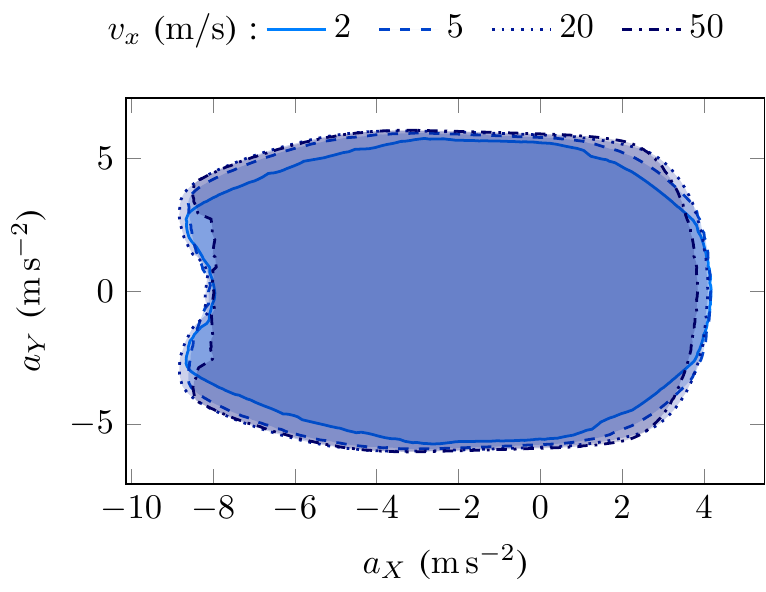}
	} 
	
	\subfloat[Yaw vs. longitudinal\label{fig:feasible-axaY}]{
		\includegtikz[width=0.9\columnwidth,height=5cm]{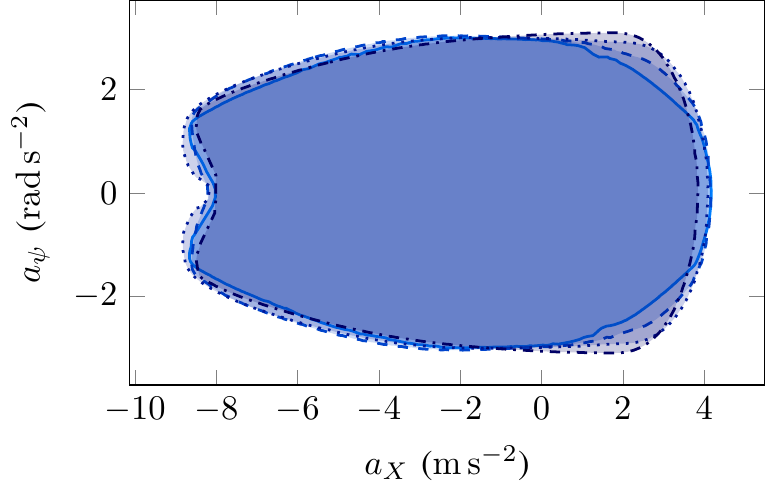}
	} 
	
	\subfloat[Yaw vs. lateral\label{fig:feasible-ayaY}]{
		\includegtikz[width=0.9\columnwidth,height=4cm]{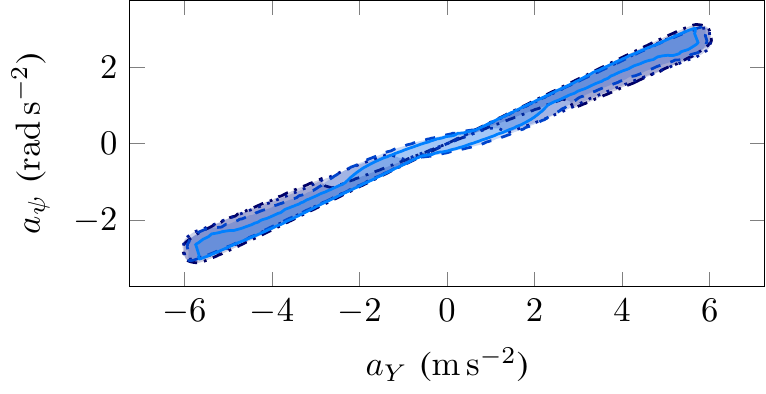}
	}
	\caption{Envelope of the computed sets of feasible accelerations for various initial longitudinal velocities $v_{x,0}$ with $v_{y,0} = 0$ and $10^5$ sampling points\label{fig:feasible}.}
\end{figure}

In \Cref{fig:feasible}, we present the computed shapes of the set of reachable accelerations in the $(a_X, a_Y)$, $(a_X, a_\psi)$ and $(a_Y, a_\psi)$ planes for a standard berline car ($l_f = \SI{1.17}{\meter}$, $l_r = \SI{1.77}{\meter}$, $l_w = \SI{0.81}{\meter}$, $M_T = \SI{1820}{\kilogram}$, front-wheel drive), over a horizon $T$ of \SI{0.1}{\second} and for various initial longitudinal velocities $v_{x,0}$. The initial state of the vehicle is taken with all angles and initial velocities (except the longitudinal one) equal to zero for the car body, and the wheels are initially rolling without slipping (\textit{i.e.} $\omega_i = v_{x,0}/r_w$ for all $i = 1\dots 4$). The friction coefficient for the road-tire contact is chosen equal to $1$. Note that this technique assumes a constant control over a time interval of \SI{0.1}{\second}; therefore, the impact of ABS or ESP cannot be measured, and may be the cause of the concavity at maximum braking observed in \Cref{fig:feasible-axay,fig:feasible-axaY} at higher velocities.

Remarkably, the projections of this set on the $(a_X, a_Y)$ and $(a_X, a_\psi)$ planes remain very similar throughout the whole speed range; namely. In the $(a_Y, a_\psi)$ plane (\Cref{fig:feasible-ayaY}), the projections are all located along the same line, except for high lateral accelerations at high speed in which over- and understeering can occur. We will use these properties to derive efficient bounds in the next section.

\subsection{Lateral velocity}
\begin{figure}[h!]
	\subfloat[Lateral vs. longitudinal\label{fig:feasible-axay-vy}]{
		\includegtikz[width=0.9\columnwidth,height=6cm]{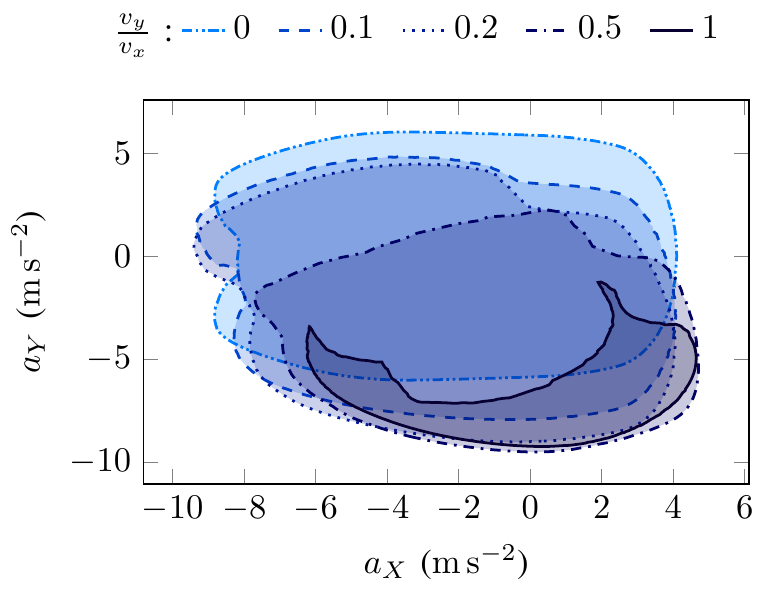}
	} 
	
	\subfloat[Yaw vs. longitudinal\label{fig:feasible-axaY-vy}]{
		\includegtikz[width=0.9\columnwidth,height=5cm]{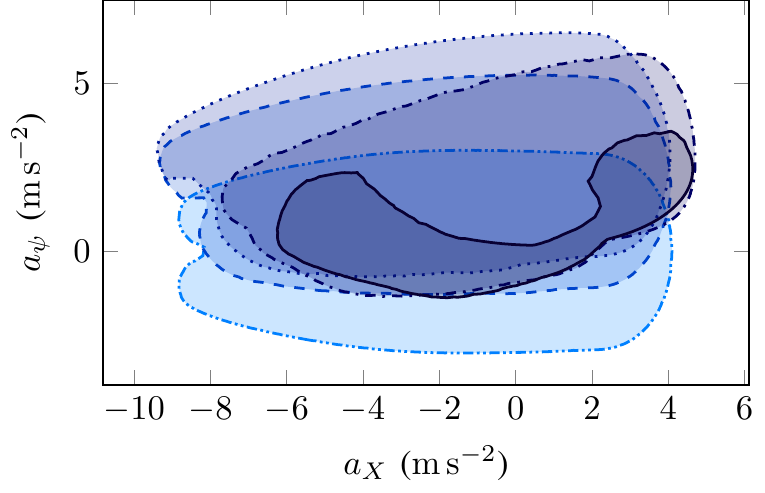}
	} 
	
	\subfloat[Yaw vs. lateral\label{fig:feasible-ayaY-vy}]{
		\includegtikz[width=0.9\columnwidth,height=4cm]{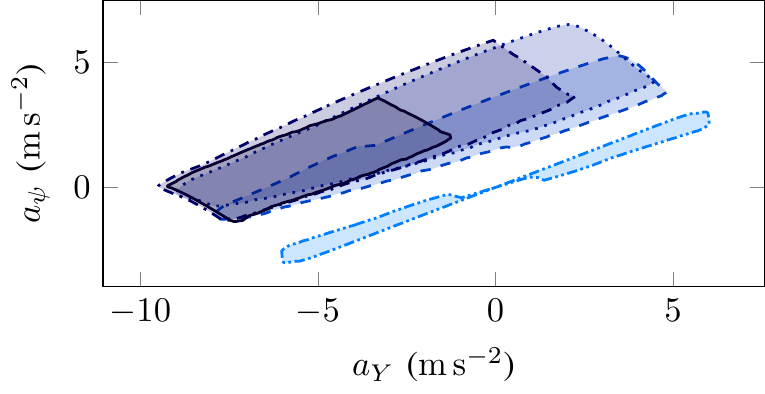}
	}
	\caption{Envelope of the computed sets of feasible accelerations for various lateral velocities $v_{y,0}$ with $v_{x,0} = $ \SI{20}{\meter\per\second} and $10^5$ sampling points\label{fig:feasible-vy}.}
\end{figure}

In \Cref{fig:feasible-vy}, we present similarly computed shapes for the sets of reachable accelerations when varying the initial lateral velocity $v_{y,0}$ with an initial longitudinal velocity $v_{x,0} =$ \SI{20}{\meter\per\second}. Initial wheel velocities are chosen, as before, as $\omega_i = v_{x,0}/r_w$ for all $i = 1\dots 4$ and initial angles and angular velocities for the car body are chosen as $0$. 

As the initial lateral velocity increases, the sets in the $(a_X, a_Y)$ and $(a_X, a_\psi)$ planes is shifted mostly along the $a_Y$ and $a_\psi$ axes respectively. Note that we also observe a slight gain in longitudinal acceleration, which corresponds to the fact that part of the initial lateral velocity can be ``redirected'' into longitudinal velocity by turning the vehicle, though this gain is very marginal. Moreover, the sets are progressively skewed as the lateral velocity increases. Interestingly, we note that increasing lateral velocity further than $0.2{v_x}$ does not provide additional acceleration performance, and instead reduces the commandability of the vehicle thus motivating to avoid these regions during planning.

\subsection{Friction coefficient}
\begin{figure}[h!]
	\subfloat[Lateral vs. longitudinal\label{fig:feasible-axay-mu}]{
		\includegtikz[width=0.9\columnwidth,height=6cm]{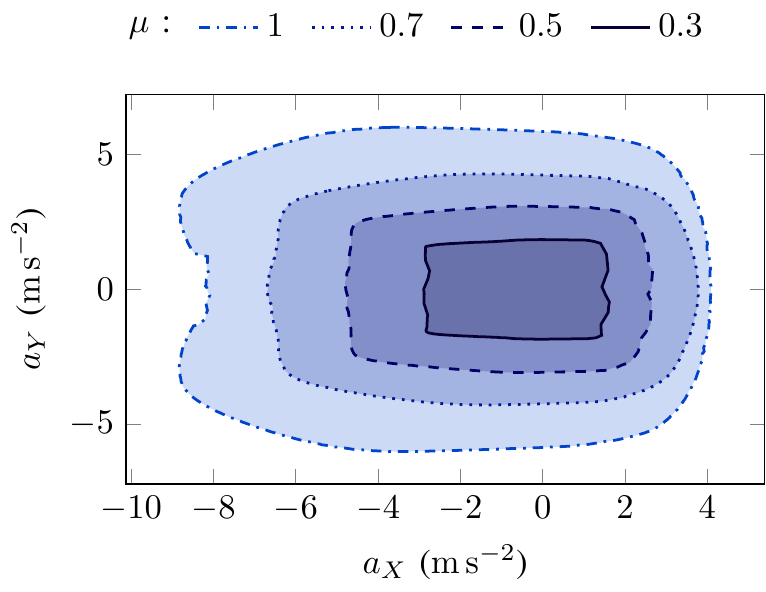}
	} 
	
	\subfloat[Yaw vs. longitudinal\label{fig:feasible-axaY-mu}]{
		\includegtikz[width=0.9\columnwidth,height=5cm]{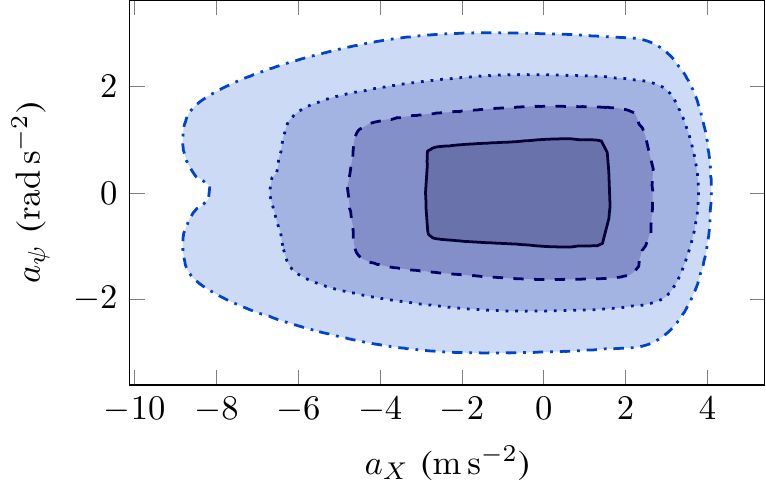}
	} 
	
	\subfloat[Yaw vs. lateral\label{fig:feasible-ayaY-mu}]{
		\includegtikz[width=0.9\columnwidth,height=4cm]{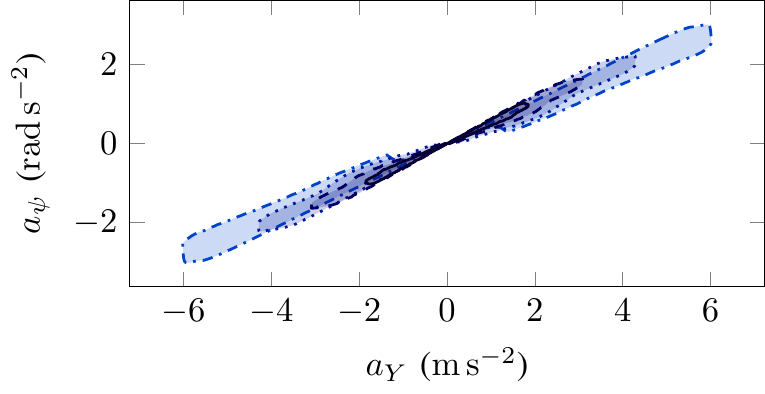}
	}
	\caption{Envelope of the sets of feasible accelerations for $v_0 = \SI{20}{\meter\per\second}$ and varying $\mu$, with $10^5$ sampling points\label{fig:feasible-mu}}
\end{figure}

In the Pacejka combined slip tire model~\cite{pacejka2005tire}, the tire-road friction coefficient $\mu$ appears both as a multiplier and a nonlinear term in the tire-road forces. In \Cref{fig:feasible-mu}, we show the variation of the envelope of feasible accelerations with $\mu$. As for the study on initial longitudinal velocity, we observe that the envelopes keep a similar shape in the $(a_X, a_Y)$ and $(a_X, a_\psi)$ planes, despite the nonlinearity of the tire model. In the $(a_X, a_\psi)$, and in spite of more important slip occurring, the reachable sets also remain aligned along the same line.

\begin{figure}
	\hspace*{-0.5cm}
	\subfloat[$\mu = 1$\label{fig:axay-mu-points10}]{
		\includegraphics[width=0.46\columnwidth,height=4cm]{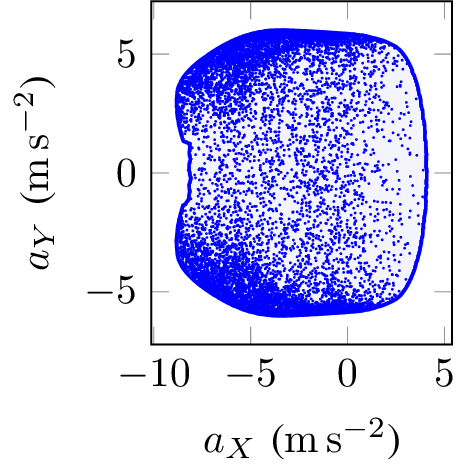}
	} 
	\subfloat[$\mu = 0.3$\label{fig:axay-mu-points03}]{
	\includegraphics[width=0.46\columnwidth,height=4cm]{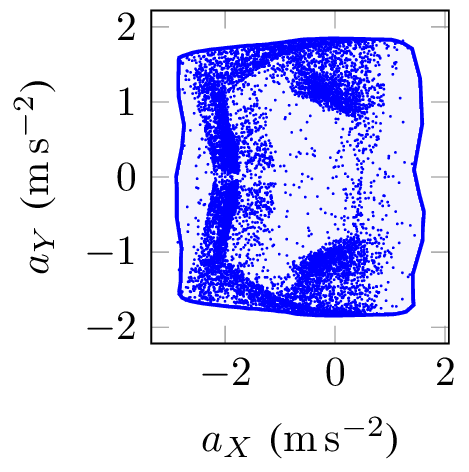}
	} 

	\caption{Accelerations corresponding to $10^4$ points (uniformly) randomly sampled in the control space, shown in in the $(a_X, a_Y)$ plane for $\mu = 1$ (dry road) and $\mu = 0.3$ (icy road). Notice the accumulation of points in \Cref{fig:axay-mu-points03}.\label{fig:sampled-mu}}
\end{figure}

Moreover, our sampling-based method evidences an interesting pattern as the friction coefficient $\mu$ decreases. \Cref{fig:sampled-mu} compares the distribution of the sampled points for $\mu = 0.3$ (icy road) and $\mu = 1$ (dry road); different scales have been chosen for better readability. For $\mu = 1$, we observe that the sampled points are almost uniformly located inside the feasible envelope. However, for $\mu = 0.3$, the sampled points accumulate near several regions of attraction, with wide areas with relatively few sampled points. This observation suggests that caution should be exercised when using direct planning methods based on sampling the control space, such as proposed in~\cite{Williams2016}, since the planner would be heavily biased towards these attraction regions, especially in low adherence situations.

\subsection{Initial rotation}
Equations \eqref{eq:globalframe} and \eqref{eq:rollpitch} show that the vehicle dynamics (except for the position in the ground coordinates) do not depend on the initial yaw angle $\psi$; therefore, the feasible regions presented above remain invariant (up to a rotation) with respect to the initial yaw angle. Moreover, we observe only very small variations of these regions for small initial values of the pitch and roll angle or rates, and variations of initial velocities of the wheels; these effects are neglected in the rest of this article.

\section{Second-order integrator model\label{sec:2d-integrator}}
Using the results from the previous section, we propose a constrained double integrator model for the vehicle dynamics. In general, such models are considered very rough approximations for the actual dynamics; however, using well-chosen constraints to couple the longitudinal, lateral and yaw accelerations, a relatively precise approximation can be obtained in this case. The proposed model considers a state vector $\bx = [X, Y, \psi, v_x, v_y, v_\psi]^T$ and a control $\bu = [u_x, u_y, u_\psi]^T$, with the same notations and reference frames as presented in \Cref{sec:vehicle-model}. The dynamic equation of the system is $\dot \bx = f_{2di}(\bx, \bu)$
with \begin{equation}f_{2di}\left(\bx, \bu \right) = \left[ \begin{array}{c}v_x \cos \psi - v_y\sin \psi \\  v_x \sin \psi +v_y \cos \psi \\ {[}v_\psi, u_x, u_y, u_\psi{]^T} \end{array} \right].\end{equation}

\begin{figure}[h!]
	\subfloat[Lateral vs. longitudinal\label{fig:feasible-axay-fullvy}]{
		\includegtikz[width=0.9\columnwidth,height=5cm]{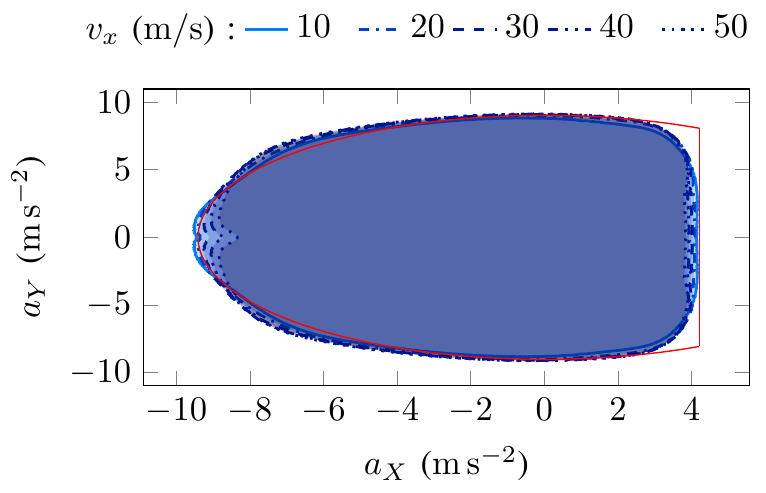}
	} 
	
	\subfloat[Yaw vs. longitudinal\label{fig:feasible-axaY-fullvy}]{
		\includegtikz[width=0.9\columnwidth,height=4cm]{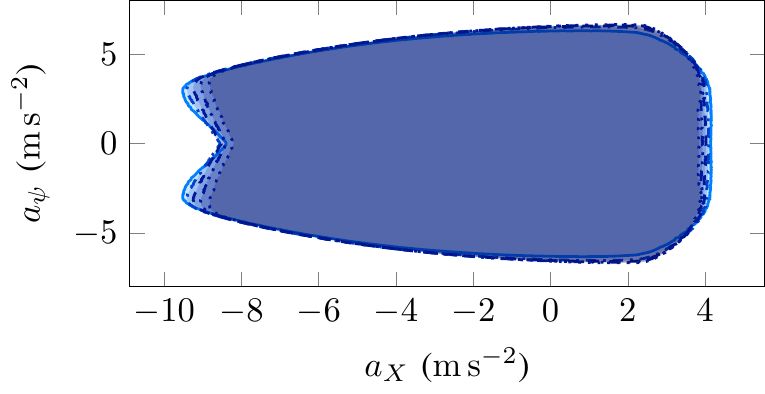}
	} 
	
	\subfloat[Yaw vs. lateral\label{fig:feasible-ayaY-fullvy}]{
		\includegtikz[width=0.9\columnwidth,height=4cm]{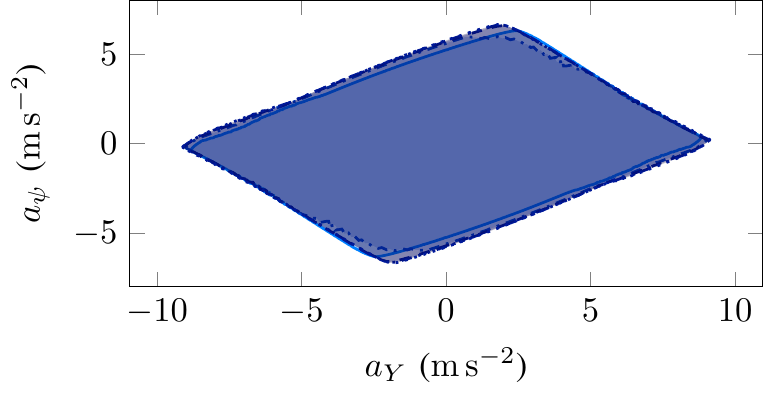}
	}
	\caption{Envelope of the full sets of feasible accelerations; notice the slight deformation along the $a_X$ axis with increasing initial velocities. The red curve in \cref{fig:feasible-axay-fullvy} shows an elliptic fit for the \SI{20}{\meter\per\second} set\label{fig:feasible-fullvy}.}
\end{figure}

In theory, it is necessary to take into account the initial state of the vehicle at each time step, and use the sets shown in \Cref{fig:feasible,fig:feasible-vy} to determine the feasible accelerations for the vehicle. However, the complex shape of these sets makes it impractical for trajectory planning. Instead, we propose to compute the ``complete'' set of feasible accelerations for the vehicle, \textit{i.e.} the union of the sets shown in \Cref{fig:feasible-vy}, which does not depend on the initial lateral velocity. These sets are shown in \Cref{fig:feasible-fullvy}; interestingly, the boundary of \Cref{fig:feasible-axay-fullvy} can be reasonably well approximated as a truncated ellipse, which is very close to the ``g-g diagram'' presented in~\cite{Funke2012}, although slightly smaller.

Using these results, we propose to approximate the sets shown in \Cref{fig:feasible-axaY} as a cropped ellipse in the $(a_X, a_Y)$ plane, and a parallelogram in the $(a_Y, a_\psi)$ plane. The parallellogram is chosen constant with the initial velocity, whereas the lower and upper bound on $a_X$ slightly var with the initial longitudinal speed. Note that a study of the 3D set of feasible accelerations (not displayed here) shows that this region is roughly convex, except for the lowest values of $a_X$. Therefore, it is only necessary to consider constraints in two of these planes to ensure that a corresponding point exists in the 3D feasible region. 


\begin{figure}
	\includegtikz[width=0.9\columnwidth,height=3.1cm]{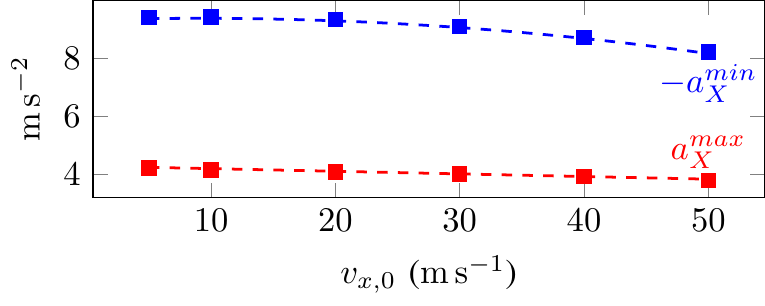}
	\caption{Variations of the $a_X^{min}$ and $a_X^{max}$ coefficients with the initial longitudinal velocity $v_{x,0}$, and polynomial fit (dashed lines). \label{fig:alpha-beta}}
\end{figure}

The resulting set of constraints on $(a_X, a_Y, a_\psi)$ can be written as:
\begin{align}
	\left( \frac{a_X}{\alpha} \right)^2 + \left( \frac{a_Y}{\beta} \right)^2 \leq 1 \\
	a_X^{min}(v_{x,0}) \leq a_X  \leq a_X^{max}(v_{x,0}) \\
	A [a_X, a_Y, a_\psi]^T  \leq b
\end{align}
where $A$ is a constant matrix, $b$ a constant vector and $a_X^{min}$, $a_x^{max}$ depend on $v_{x,0}$. For the proposed vehicle, the experimental data of \Cref{fig:feasible-fullvy} yield $\alpha =$ \SI{9.4}{\meter \per \second\squared},  $\beta =$ \SI{9.0}{\meter \per \second\squared}, $A = \left( \begin{smallmatrix} 2.6 & 1 & 0 \\ 2.6 & -1 & 0 \\ 0 & 1.1 & 1 \\ 0 & -1.1 & -1 \\ 0 & -0.57 & 1 \\ 0 & 0.57 & -1 \end{smallmatrix} \right)$ and $b = \left( \begin{smallmatrix} 15.3 \\ 15.3 \\ 9.9 \\ 9.9 \\ 5.1 \\ 5.1 \end{smallmatrix} \right)$ \SI{}{\meter\per\second\squared}. The evolution of $a_X^{min}$ and $a_X^{max}$ with $v_{x,0}$ are shown in \Cref{fig:alpha-beta}; a polynomial fit yields $a_X^{min}(v_{x,0}) = -9.3 - 0.013v_{x,0} + 0.00072{v_{x,0}}^2$ and $a_X^{max}(v_{x,0}) = 4.3 - 0.009v_{x,0}$ (with $v_{x,0}$ expressed in \SI{}{\meter\per\second} and accelerations in \SI{}{\meter\per\second\squared}).

Note that these constraints only guarantee the feasibility of a trajectory. To actually drive the vehicle, it is necessary to find a high-frequency low-level control loop capable of following this feasible trajectory. Moreover, the current set of constraints does not account for limitations on the actuator dynamics, which will be the subject of future work.

\section{Numerical results}\label{sec:numerical-sim}
The previous section provides a set of conditions for the dynamic feasibility of a trajectory, in the form of bounds on vehicle acceleration. These constraints can be used to design a trajectory planner for the vehicle, for instance using model predictive control (MPC). The specifics of our MPC implementation are out of the scope of this paper, and are presented in details in \cite{Altche2017}. In this section, we simply present figures extracted from \cite{Altche2017} to illustrate the good performance of our model when compared to a more classical kinematic bicycle one~\cite{Kong2015}. For both models, the planner is used to drive a vehicle along a circuit at high speed; the planning horizon is chosen as $T = \SI{3}{\second}$, and the time step of the MPC solver (based on the ACADO Toolkit~\cite{Houska2011}) is \SI{0.2}{\second}.


\begin{table}
	\caption{Synthesis of performance for both planners\label{tab:comput-time}}\centering
	\begin{tabular}{l c c c}
		\toprule
		Model & Avg. comp. time & RMS lat. error & Max. lat. error \\
		\midrule
		Proposed & \SI{59.8}{\milli\second} & \SI{0.25}{\meter} & \SI{0.70}{\meter} \\
		Kinematic&  \SI{57.8}{\milli\second} & \SI{0.16}{\meter} &  \SI{0.95}{\meter} \\
		\bottomrule
	\end{tabular}
\end{table}

\begin{figure}
	\includegtikz[width=0.9\columnwidth,height=4cm]{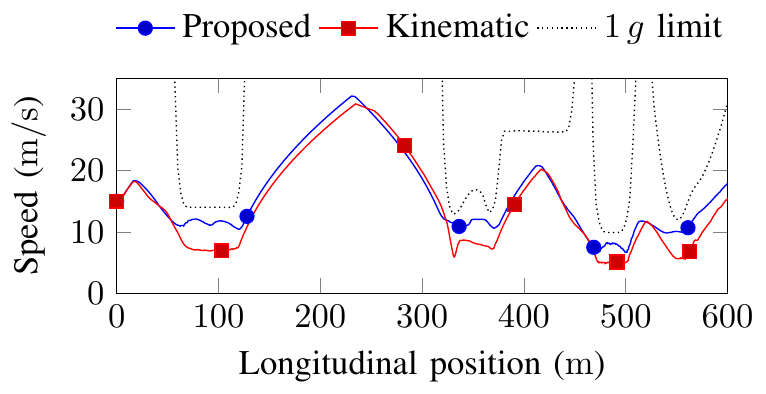}
	\caption{Comparison of achieved speed for both planners, alongside with a theoretical maximum corresponding to $1\, g$ lateral acceleration in curves. \label{fig:comp-speed}}
\end{figure}

\begin{figure}
	\includegtikz[width=0.9\columnwidth,height=3.5cm]{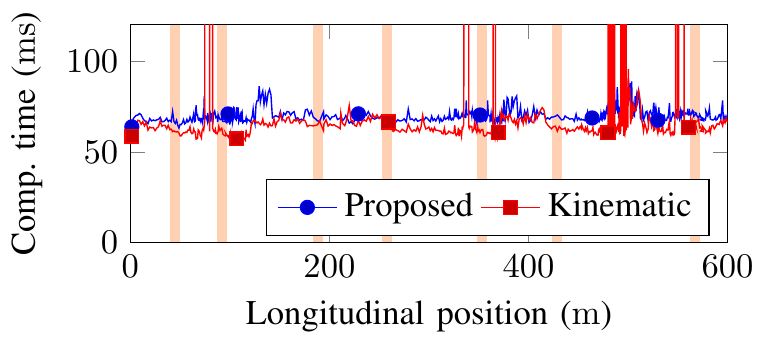}
	\caption{MPC computation time (with obstacles) for the proposed model, and for a kinematic bicycle model starting from the same state. \label{fig:comput-time}}
\end{figure}

\Cref{tab:comput-time} provides synthetic performance data for both planners when driving around a circuit, showing roughly similar computation time and lateral error. However, \Cref{fig:comp-speed} shows that our planner achieves significantly higher speeds in curves, nearing lateral accelerations of $1g$ while reducing the maximum lateral error. Moreover, an interesting property of our proposed model is that it makes the corresponding MPC planner more robust: in presence of obstacles, the kinematic planner sometimes fails to output a solution in real-time (as shown in \Cref{fig:comput-time}); this phenomenon does not occur with our proposed model. In both cases, the superior performance obtained with the proposed model is likely due to the simpler relations between the outputs (future positions) and the control inputs, which allows the solver to converge faster, towards better solutions. 

\section{Conclusion\label{sec:conclusion}}
In this article, we proposed a new modeling of vehicle dynamics as a constrained second-order integrator. First, we described a high fidelity 9 degrees of freedom vehicle model including tire slip and load transfer. We used this model with an offline random sampling technique to show that the proposed second-order model, despite its simplicity, is able to capture most of the relevant dynamics of the vehicle up to its handling limits. Moreover, we showed that this model is also compatible for driving on slippery roads, at the cost of a changing a few parameters. Implementation of the second-order integrator model inside an MPC-based trajectory planner shows that computation time remains roughly similar compared to using a kinematic bicycle model, but solution quality and robustness seem to be improved, notably in presence of obstacles. These results open several research perspectives since planning aggressive trajectories has often been thought to necessitate highly precise, and thus complex, vehicle models. Future research will focus on designing a low-level control law capable of precisely tracking the generated feasible trajectories, and validating these results on a real, scale model of a vehicle.

\bibliographystyle{IEEEtranurl}
\bibliography{dynamic}

\end{document}